# REAL TIME PARAMETER ESTIMATION FOR ADAPTIVE OFDM/OTFS SELECTION

Amina Darghouthi[1], Abdelhakim Khlifi[2], Belgacem Chibani[1]

[1]MACS Laboratory: Modelling, Analysis and Control of Systems, University of Gabes, Tunisia.
[2]Innov'COM laboratory, Sup'COM, University of Carthage, Tunisia.

## ABSTRACT

*Future wireless communication systems must simultaneously address multiple challenges to ensure accurate data detection, deliver high Quality of Service (QoS), adding enable a high data transmission with low system design. Additionally, they need to reduce energy consumption and latency without increasing system complexity. Orthogonal Frequency Division Multiplexing (OFDM) is a commonly used waveform in 4G and 5G systems, it has limitations in handling significant delay and Doppler spread in high mobility scenarios. To overcome these weaknesses, a novel waveform named Orthogonal Time Frequency Space (OTFS) has been proposed, which aims to improve upon OFDM by closely matching signals to channel behavior. In this study, we propose a novel strategy that enables operators to dynamically select the best waveform based on estimated mobile user parameters. We use an Integrated Radar Sensing and Communication System (ISAC) to estimate delay and Doppler, as well as speed and range. This approach allows the base station to adapt to the mobile target, thereby enhancing the performance of wireless communication systems in high mobility and low complexity scenarios. Simulation results demonstrate the effectiveness of our proposed approach and show that it outperforms existing methods.*

## 1. INTRODUCTION

Wireless communication systems are continuously evolving. They lookformeeting the increasing demands of various mobility situations. Orthogonal Frequency Division Multiplexing (OFDM) is a widely used technology. Unfortunately, it faces significant challenges in high mobility environments. This include Doppler shifts and spread effects [1], [3], [4], [7], [21]. In order to address these limitations, a new processing approach has been developed that defines a new modulation technique called Orthogonal Time Frequency Space (OTFS) [1], [7]. By leveraging delay and Doppler diversity, OTFS outperforms OFDM in high mobility scenarios, making it a promising candidate for High Mobility Wireless Communication systems (HMWCS) [1], [2], [3].Besides, accurately estimating delay Doppler is a significant challenge for future wireless systems, including 6G.Even though, 4G/5G technologies have made improvements in mobility scenarios, interruptions can still occur at very high speeds. To deal with this issue, 5G networks must be capable of supporting speeds up to 500km/h in applications such as Vehicle to Everything (V2X), drones, and HighSpeed Rails (HSR) [1], [2], [5], [6]. Recently, Multi-Carrier Modulations (MCM) have been under consideration to tackle the challenges posed by frequency selective channels, which involve conducting actions in the frequency domain. With the advent of high mobility scenarios such as hyper loop, it is anticipated that future 6G networks will need to support mobility at speeds up to 1000km/h [1], [5]. Nevertheless, such high mobility introduces significant Doppler shift and spread, resulting in HMWCexperiencing rapid selective double fading. In order to effectively communicate in such conditions, it is crucial to match the

                                                                                   109

information to the propagation channel.The clever utilization of cyclic redundancy during transmission allows for the reduction of terminal complexity thanks to FFT (Fast Fourier Transform) usages based algorithms. This waveform suffers from some limitations, including a high Peak to Average Power Ratio (PAPR), Out of Band (OOB) emissions, and a significant loss of orthogonal waves in high mobility [7].So, it is necessary to explore alternative modulation techniques that can overcome these limitations and provide reliable communication in higher mobility situations. Recently a proposed bi-dimensional waveform, known as OTFS (Orthogonal Time Frequency Space), utilizes a pair of 2D transforms, namely Symplectic Finite Fourier transform (SFFT) and Inverse Symplectic Finite Fourier transform (ISFFT) [1], [8], [9], [15]. OTFS has been shown to achieve full diversity and outperform OFDM in high mobility scenarios, making it a promising candidate for the next generation of mobile networks [11], [12], [13], [16], [17]. Both OTFS and OFDM waveforms have their own pros and cons. Particularly, OTFS excelles in high mobility scenarios but it suffers from high processing complexity.While OFDM is well suited for low mobility scenarios but it experiences significant performance degradation in higher mobility cases [1], [12].To address these drawbacks.This paper suggests an original idea to explore an alternative usage of these waveforms. This guaranteesa high quality of service (QoS) simultaneously for different mobility contexts. One promising approach is the use of an Integrated Sensing and Communication (ISAC) system for estimating various parameters, particularly the delay(range) and Doppler (speedof moving objects), to implement OTFS and OFDM schemes accurately. The objective is to achieve more accurate estimates of delay, Doppler shift, object velocity, and target count. Traditional velocity estimation methods often rely on the delay Doppler (DD) technique, but the proposed approach suggests a novel solution to enhance the performance of wireless communication systems in various mobility scenarios. Recently, the integration of radar algorithms for parameter estimation has gained significant attention in the research community [20], [22], [23], [24], [25]. In wireless communication systems, users are distributed randomly within the coverage area of the base station and exhibit varying levels of mobility [31], [32].Therefore, it is crucial to propose effective solutions that cater to the needs of both faster and slower users simultaneously when dealing with a diverse group of users with varying mobility levels within the base station's coverage area.

## 2. RELATED WORK

The advent of 6G technology brings forth a myriad of challenges and prospects for research. One of the primary focuses is the examination of various parameters employed for waveform sensing in Integrated Sensing and Communication (ISAC) systems [22], [23], [24], [25]. This research area aims to identify the most suitable waveform parameters for sensing signals and optimize the allocation of bandwidth between sensing and communication [22], [23], [24], [25]. Researchers can unlock new opportunities for enhancing network capabilities and enabling advanced applications, such as autonomous driving and industrial automation, by tackling the challenges faced by future wireless communication systems.The importance of accurate delay and Doppler shift parameter estimation in Integrated Sensing and Communication Systems (ISAC) is highlighted in [18], [19]. To address this challenge, the author of [19] proposes a novel two-stage estimation algorithm, the Fibonacci matched filter (MF-F). This algorithm effectively utilizes waveform characteristics in the Doppler delay shift (DD) domain within orthogonal time frequency space (OTFS), enhancing the overall performance of the ISAC system. Moreover, the studies[23], [26] explore potential research areas, including wireless propagation path prediction, electromagnetic spectrum mapping, and Terahertz technology. To address these domains, the authors propose an Integrated Sensing and Communication (ISAC) estimation system capable of accurately estimating delay, Doppler shifts, velocity, and range of moving objects. The studies [18] and [19] emphasize the significance of accurate delay and Doppler shift parameter estimation in Integrated Sensing and Communication Systems (ISAC). To address this challenge, the author of [19] proposes a novel two steps estimation algorithm, the Fibonacci-matched filter (MF-F).This algorithm effectively utilizes waveform characteristics in the Doppler delay shift (DD) domain within orthogonal time-frequency space (OTFS), enhancing the overall performance of the ISAC system. Furthermore, the studies[23], [26] explore potential research areas, including wireless propagation path prediction, electromagnetic spectrum mapping, and Terahertz technology. To address these challenges, the authors propose an Integrated Sensing and Communication (ISAC) estimation system capable of accurately estimating delay, Doppler shifts, velocity, and range of moving objects. Besides, the study in [27] showcases the potential of simultaneously estimating range, Doppler, and azimuth information for any number of objects during transmission, relative to the number of array antenna elements. Among different algorithms used for estimating channel conditions (CE) based on antenna data, MUSIC has been extensively utilized. The implementation of the MUSIC algorithm is straightforward, and it has been modified in various versions to cater to different scenarios while providing high resolution [27]. In addition, the article [28] presents a novel integration strategy for moving vehicle detection and wireless communications utilizing OTFS modulation. By harnessing channel impulse response coefficients, this approach achieves precise detection of mobile targets during wireless transmission, estimating distances and speeds of reflecting targets. Furthermore, incorporating random pilots in OTFS modulation improves vehicle speed estimation in noisy environments, providing a more reliable solution for moving target detection than conventional methods.According to [29], the authors introduce new algorithms that are being developed, enabling the joint estimation of target parameters like angle, range, and velocity using MIMO-OFDM waveforms, ultimately enhancing the performance of ISAC systems. Authors in [30] have also, propose a new method for creating a lower complexity automotive radar system using a narrowband OTFS ISAC (Integrated Sensing and Communication) system. By utilizing a single OTFS support as a pilot, this method enables accurate estimation of fractional delay and Doppler shifts, providing higher resolution range velocity profiles. Furthermore, authors in [32] presents an Integrated Sensing and Communication (ISAC) system using Orthogonal Time Frequency Space (OTFS) modulation to obtain accurate range velocity profiles. This system uses an OTFS support with a rectangular pulse as a pilot to simultaneously estimate delay and Doppler shift, enabling the determination of the range and velocity of radar targets.Table 1 below givesa comprehensive summary of the various radar techniques utilized in automotive applications, along with their individual contributions and complexity levels.The Table compares a range of radar techniques, including object velocity estimation, position estimation, and parameter estimation. Furthermore, Table 1 introduces a new technique aimed at improving the estimation of delay and Doppler shift parameters, thereby facilitating a smooth transition between OFDM and OTFS. The methods employed in this technique encompass Fast Fourier Transform (FFT) for OFDM, the Fibonacci matched filter (MF-F) algorithm for OTFS, and Symplectic Fast Fourier Transform (SFFT).

Table 1. Literature on Estimating Algorithms.

| Ref | Contributions | Methods used | Complexity |
| --- | --- | --- | --- |
| [19] | Aims to enhance the estimation of delay parameters and Doppler shift. | MF-Fibonaccimatched filter | Medium |
| [20] | Focuses on estimating the speed of objects using MUSIC algorithm. It enables simultaneous estimation of range, Doppler, and azimuth information. | MUSIC algorithm, Radar beam forming techniques | Low Medium |
| [28] | A novel estimation method for processing MIMO-OFDM communication waveform echoes, which jointly estimates target angle-range-velocity parameters. Also provides theoretical analysis for the maximum unambiguous range, resolution. | MIMO OFDM radar | Low Medium |
| [31] | An OTFS-based ISAC system for automotive radar applications, enabling high-resolution range-velocity profiles. | Fourier transform's self-duality and ability to maintain shape under time delay and Doppler | Low Medium |
| Proposed | Enhancement of delay and Doppler shift parameter estimation for a switch between OFDM and OTFS. The methods employed include FFT for OFDM, the SFFT and the MF-F algorithm for OTFS. | For OFDM: FFT For OTFS: SFFT and MF-F | Low |

### A. CONTRIBUTIONS

This paper presents, a novel switching strategy that dynamically selects the optimal waveform for mobile users by integrating radar sensing and communication in a hybrid OTFS-OFDM system. This approach uses mobile parameter estimation, facilitated by the ISAC radar and the Matched Filter Fast Fourier (MF-F) algorithm, to improve accuracy and efficiency.The system's performance is enhanced by choosing between OFDM or OTFS strategies based on estimated user data detection parameters, with delay and Doppler values being crucial in the selection process. This methodology aims to improve QoS, reduce complexity, and ensure accurate data detection in high mobility and low complexity scenarios, outperforming existing methods and showing potential for future 6G networks.Based the aforementioned paper, this manuscript will present the following key contributions:

- The suggested system model attains a high range resolution for switching between OFDM/OTFS based on mobile parameter estimation. By incorporating ISAC radar sensing into the system model, we are able to achieve more accurate parameter estimates. This approach effectively reduces the cost and complexity and power consumption associated with implementing such systems.

- The proposed estimator provides the lowest variance estimate among unbiased estimators and achieves the lowest possible Root Mean Squared Error (RMSE) compared to all unbiased methods.
- Finally, the evaluation of error probability in the proposed system demonstrates that OTFS outperforms OFDM in high mobility scenarios. Additionally, the system can be further improved by setting threshold values for parameter selection to optimize OFDM performance over OTFS. The simulation results align well with the expected outcomes from theoretical models.

**B. ORGANISATION AND NOTATION**

Such pros will be highlighted, over this proposed paper structured as follows: Section 2 provides an overview of OFDM and OTFS systems, Section 3 introduces the proposed system model and parameter estimation methods, Section 4 presents simulation results and discusses system complexity, and finally, Section 5 concludes the paper with a summary of findings and suggestions for future research.

The circular convolution operator $\circledast$; The complex conjugate of the transmitter and the received signal are denoted by the symbol $g_{tx}^*$ and $g_{rx}^*$.

## 3. SENSING SYSTEM MODEL

### 3.1. OFDM signal Modelling

In OFDM transmission, data is transmitted through narrow subcarriers within the bandwidth, where each subcarrier sends M-QAM symbols to an OFDM modulator. Although the transmission is successful over the channel, Inter Symbol Interference (ISI) is a common occurrence. The modulation and demodulation tasks can be efficiently performed using Fast Fourier Transform (FFT) and its inverse (IFFT), as illustrated in Fig.1.

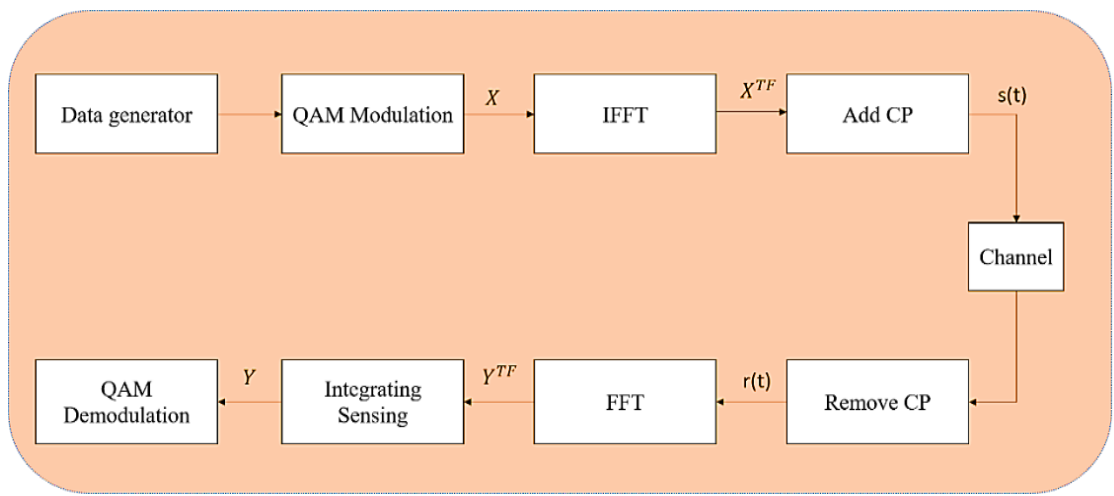

Fig1. OFDM Transmitter and Receiver Block Diagram

To mitigate the issue of Inter Symbol Interference (ISI) in OFDM transmission, a cyclic prefix (CP) is inserted between consecutive OFDM symbols. The CP length should be greater than the channel delay spread to effectively reduce ISI and simplify the equalization process [1], [10], [11], [13]. In an OFDM system with M subcarriers and N time slots, the total bandwidth of the International Journal of Computer Networks & Communications (IJCNC) Vol.16, No.4, July 2024OFDM signal is $B = M \Delta f$, where $\Delta f$ is the subcarrier spacing set to $\frac{1}{T}$. The frame duration is $T_f = NT = NMT_s$, with T representing one OFDM symbol duration equal to $MT_s$, where $T_s$ is the sampling time. Assuming a static multipath channel with a maximum delay spread of $\tau_{max}$, the channel delay is caused by $\frac{\tau_{max}}{T_s}$. To address ISI and simplify channel equalization, the cyclic prefix length $L_{CP}$ should be greater than or equal to $l_{max}$. In this scenario, $L_{CP}$ is chosen to be equal to $l_{max}$. The data symbols are defined as [11], [12], [14], [18]:

$$X[m,n] = m = 0, \dots, M-1; n = 0, \dots N-1 \quad (1)$$

The data symbols are extracted from the alphabet $A = \{a_1, \dots, a_Q\}$, where $Q$ is the number of unique symbols in the alphabet. Each column of the matrix X contains N symbols. The transmitted signal $s(t)$ can be represented as [11], [12], [14], [18]:

$$s(t) = \sum_{n=0}^{N-1}\sum_{m=0}^{M-1} X[m,n] g_{tx}(t-nT) e^{j2\pi m \Delta f(t-nT)} \quad (2)$$

The pulse shaping waveform $g_{tx}(t)$ is used to model the continuous time transmitted signal, where $g_{tx}(t) \geq 0$ for $0 \leq t < T$. The collection of orthogonal basis functions $\phi_{(n,m)}(t)$ is established to shape the M-QAM symbols as follows [11], [12], [14], [18]:

$$\phi_{(n,m)}(t) = g_{tx}(t-nT) e^{j2\pi m \Delta f(t-nT)}, {}_{0 \leq m \leq M; 0 \leq n \leq N} \quad (3)$$

The received basic signals are utilized by the receiver in order to multiplex information, as described in [10], [12], [13].

$$\phi_{(n,m)}(t) = g_{rx}(t-nT) e^{j2\pi m \Delta f(t-nT)}, {}_{0 \leq m \leq M; 0 \leq n \leq N} \quad (4)$$

where, $g_{rx}(t) \geq 0$ for $0 \leq t < T$ and is zero otherwise. Equation 5, can rewritten as following [10], [12], [13]:

$$s(t) = \sum_{n=0}^{N-1}\sum_{m=0}^{M-1} X[m,n] \phi_{(n,m)}(t) \quad (5)$$

After that, a CP extension is then added to signal $s(t)$ in order to overcome multipath channel's effects. The cross-ambiguity function between the two signals $g_1(t)$ and $g_2(t)$ are defined [10], [12], [13]:

$$A_{g_1,g_2}(f,t) \triangleq \int g_1(t) g_s^*(t'-t) e^{-j2\pi f(t'-t)} dt' \quad (6)$$

The ambiguity function $A_{g_1,g_2}(f,t)$ defines the correlation between $g_1(t)$ and $g_2(t)$ functions delayed by t and shifted in frequency by f and t in the time-frequency plane. When the transmitted signal $s(t)$ passes through a time and frequency-selective radio channel, the received signal in the time domain is known as $r(t)$.

Let $r(t)$ be the received time domain signal after propagation through a time-frequency selective wireless channel. This channel is characterized by its impulse response $h(t)$. So, the received signal can be expressed as [10], [12], [13]:

$$r(t) = h(t) \circledast s(t) + w(t) \qquad (7)$$

Received signal for time domain, $r(t)$ is obtained after the transmitted signal, $s(t)$ passes through a time-frequency selective wireless channel characterized by its impulse response, $h(t)$. The received signal is given by the convolution of the transmitted signal with the channel impulse response, along with the addition of a centered complex Gaussian white noise, $w(t)$. After the removal of cyclic prefix (CP), the time-frequency samples received, $r(t)$ is projected onto each orthogonal basis function, $\phi_{(n,m)}(t)$. It can be expressed as an FFT operator applied to continuous time, as stated in [10], [12], [13]:

$$Y(f,t) = A_{g_1,g_2}(f,t) \triangleq \int g_{rx}^*(t'-t)e^{-j2\pi m\Delta f(t'-t)}, \qquad (8)$$

$$Y(m,n) = Y(f,t)_{f=m\Delta f, t=nT}. \qquad (9)$$

## 3.2. OTFS System Modelling

In this section, we will focus on the well-known concept introduced by Hadani, referred to as the OTFS approach [7], [15]. Particularly, we will present the system model associated with the OTFS scheme, as depicted in Fig.2.

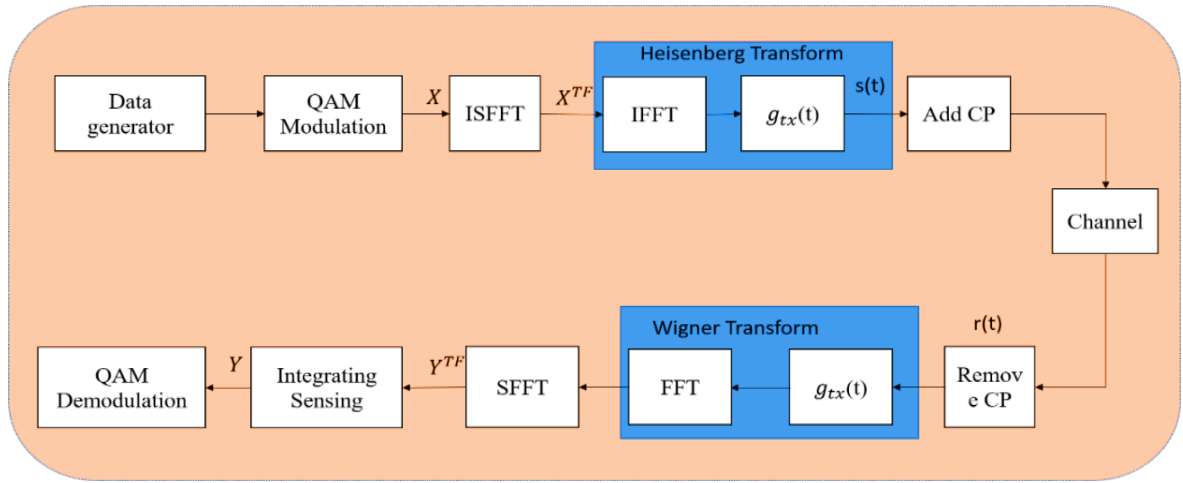

Fig.2. Block Diagrams of OTFS system

The OTFS technique involves the mapping of prepared QAM symbols onto the delay-Doppler domain (DD). This mapping is achieved by initially converting the symbols from the delay-Doppler domain to the time-frequency domain (TF) at the transmitter. The QAM symbols are arranged in a two-dimensional matrix with $N$ columns in the Doppler domain and M rows in the delay domain. The time-frequency grid is discretized into a $M$ by $N$ grid, using intervals of $T(s)$ and $\Delta f\ (Hz)$, to sample the time and frequency axes, respectively. This discretization is performed for some integers $N, M > 0$, as demonstrated in [11], [12], [18], [31].

$$\Lambda = \{(nT, m\Delta f)_{n=0,\cdots,N-1, m=0,\cdots,M-1}\} \qquad (10)$$

The modulated time frequency samples $X[n,m], n = 0, \ldots, N-1, m = 0, \ldots, M-1$ are transmitted over an OTFS frame with a duration of $T_f = NT$ and a bandwidth of $B = M\Delta f$.

The delay Doppler plane $\Gamma$ in the region $(0,T] \times (\frac{-\Delta f}{2}, \frac{\Delta f}{2})$ is discretized into an $M$ by $N$ grid, as shown in[11], [12], [18], [31].

$$\Gamma = \{(\frac{k}{NT}, \frac{l}{M\Delta f}), k = 0,\cdots,N-1, l = 0,\cdots,M-1\} \quad (11)$$

where $\frac{k}{NT}$ and $\frac{l}{M\Delta f}$ represent the quantization steps of the delay and Doppler frequency axes, respectively.

In the next step, the Inverse Symplectic Finite Fourier Transform (ISFFT) is applied to the signal to convert it from the DD domain to TF domain. This transformation can be represented as follows [11], [12], [18], [31].

$$X^{TF}[m,n] = \frac{1}{\sqrt{MN}} \sum_{n=0}^{N-1} \sum_{m=0}^{M-1} X[l,k] e^{j2\pi(\frac{nk}{N} - \frac{ml}{M})} \quad (12)$$

where $X[l,k]$ is the delay Doppler signal modulated pulse.

In this scenario, each data frame has a total frame duration of $B = NT$ and a bandwidth of $T_s = N\Delta f$. The matrix $X^{TF}[m,n]$ is reshaped into a time-frequency domain sequence, and the transmitted OTFS signal, denoted as $s(t)$, is obtained by applying the Heisenberg transform to $X^{TF}$ with the transmitter shaping pulse, $g_{tx}(t)$.

The Heisenberg transform is a form of multicarrier modulation that utilizes the traditional OFDM modulator. Specifically, the Heisenberg transform can be implemented through an inverse fast Fourier transform (IFFT) module and transmit pulse shaping within the framework of conventional OFDM modulation. As explained in [11], [12], [18], [31], the signal $s(t)$ is transmitted in the time domain using the Heisenberg Transform.

$$s(t) = \sum_{n=0}^{N-1} \sum_{m=0}^{M-1} X^{TF}[m,n] g_{tx}(t-nT) e^{j2\pi m\Delta f(t-nT)} \quad (13)$$

where, $g_{tx}(t)$ is the window function. In practice, rectangular transmit and receive pulses that are compatible with OFDM modulation are utilized. Lastly, a cyclic prefix (CP) is added to the time domain signal for each data frame, as suggested in [11], [12], [18], [31], [32].

$$S_{CP}(t) = \begin{cases} s(t) & 0 \leq t \leq T_s \\ s(t+T_s) & -T_{CP} \leq t < 0 \end{cases} \quad (14)$$

where $T_{CP}$ denotes the duration of the CP.

### 3.3. Channel Model and Received Signal

The impulse response of the wireless channel in the delay Doppler (DD) domain is characterized by the target detection channel or communication paths transmitted. We assume that there are multipath components (P), where the i[th] path is associated with a complex path gain $\alpha_i$, delay $\tau_i$, and Doppler shift $v_i$. In this scenario, any two paths are resolvable in the delay Doppler domain

International Journal of Computer Networks & Communications (IJCNC) Vol.16, No.4, July 2024

(i.e., $|\tau_i-\tau_j| \geq \frac{1}{M\Delta f}$ or $|v_i - v_j| \geq \frac{1}{NT}$). Therefore, the impulse response of the wireless channel in the DD domain is given as [30], [31], [32]:

$$h(\tau, v) = \sum_{i=0}^{P} \alpha_i \, \delta(\tau - \tau_i)\delta(v - v_i) \quad (15)$$

In order to develop a radar ISAC system that integrates both sensing and communication capabilities, it is essential to calculate the delay and Doppler shifts. These shifts can be computed using the equations $\tau_i = \frac{r_i}{c_0}, v_i = \frac{f_c v_i}{c_0}$;

where $r_i$ and velocity $v_i$ along the i[th] path, and $f_c$ is the carrier frequency and the speed of light is therefore represented by $c_0$.

The successful integration of system detection requires careful evaluation of both the round-trip delay and the Doppler effect. In this scenario, the path delay and Doppler shift correspond to integer multiples of the delay and Doppler resolution, denoted as $\tau_i = \frac{l_i}{M\Delta f}$ and $v_i = \frac{k_i}{NT}$, respectively.

In high mobility scenarios, the transmitted signal can undergo various changes, resulting in shifts in both the time and frequency domains. Consequently, the received signal matrix $Y^{TF}[m, n]$ is obtained by sampling the crossambiguity function $A_{g_{rx,r}}(t, f)$ according to [13], [14], [16],[30],[31],[32]

$$Y^{TF}[m, n] = A_{g_{rx,r}}(t, f)_{t=nT, f=m\delta f} \quad (17)$$

where the sampling cross ambiguity function $A_{g_{rx,r}}(t, f)$ as indicated:

$$A_{g_{rx,r}}(t, f) \triangleq \int g_{rx}^*(t - t')r(t)e^{j2\pi(t-t')} \quad (18)$$

Finally, the DD domain samples are obtained by applying the SFFT to $Y[l, k]$ [13], [14], [16],[30], [31], [32]:

$$Y[l, k] = \frac{1}{\sqrt{MN}} \sum_{n=0}^{N-1} \sum_{m=0}^{M-1} X[m, n] e^{j2\pi(\frac{nk}{N} - \frac{ml}{M})} \quad (19)$$

Where $X[m, n]$ is the signal using the inverse symplectic finite Fourier transform (ISFFT).

## 4. PROPOSED ARCHITECTURE

In this section, we will present the appropriate signal processing method based on a selected approach for the analysed signal. This method pertains to the ISAC radar's technique for estimating various parameters, which is a valuable tool for identifying and estimating the parameters of unknown objects that characterize a mobile user.

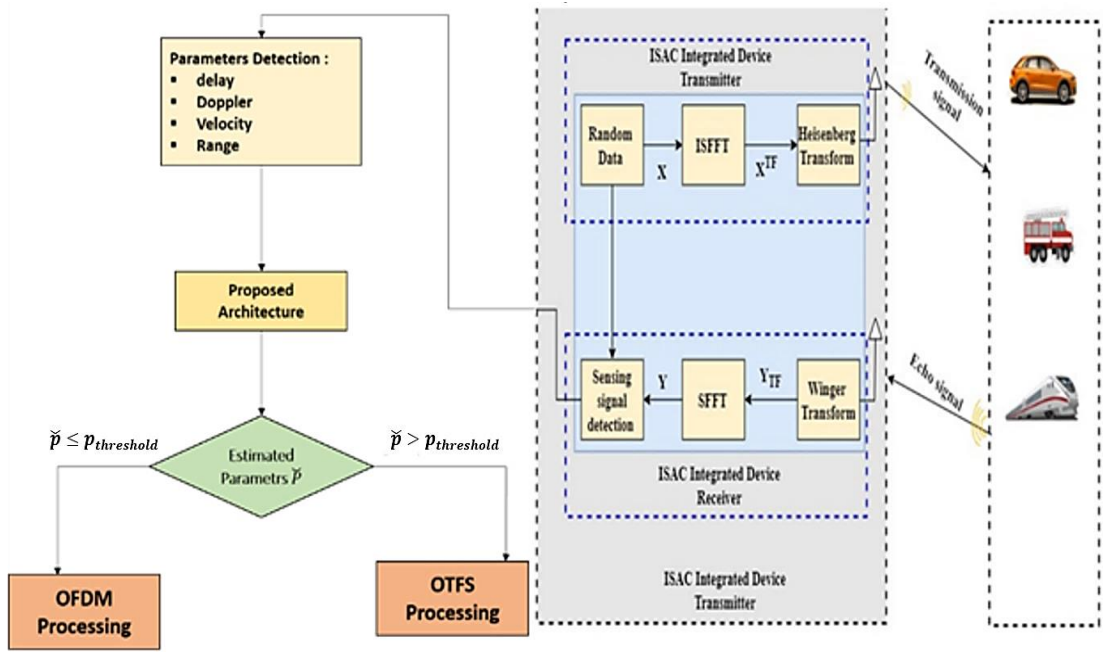

Fig.3. Proposed architecture based on a real time parameters estimation.

As illustrated in Fig.3, the framework of the signal processing system, which contains three primary processing blocks. The first block is the base station, which includes components such as Random Data, Inverse Fourier Transform Symplectic (ISFFT), and Heisenberg Transform. The base station transmits a signal to the Sensing Target, which is a mobile object such as a vehicle. The Sensing Target then returns an echo signal to the receiver in the base station, forming an integrated ISAC device. The ISAC receiver processes the signal using various tools such as the Wigner transform, Fourier Transform Symplectic (SFFT), and Sensing Signal Detection. The Sensing Signal Detection is responsible for estimating various parameters such as delay spread, Doppler frequency, velocity, and range of the detected objects. These signals are processed by comparing them to predefined thresholds for range (delay), velocity (Doppler frequency), known to as Parameter thresholds [1], [29], and [31]. These thresholds are essential in distinguishing between lower and higher mobility cases. If the estimated parameters fall below the threshold, the Sensing Signal Detection sends a signal to the base station to proceed with the OFDM signal processing tasks. Conversely, if the speed exceeds the threshold, the OTFS signal processing is retained. In summary, the OFDM approach is primarily applicable in lower mobility. The previous analysis has led us to propose an architecture for a hybrid OTFS-OFDM system that utilizes user estimation based on the ISAC radar application's approach for estimating various parameters. Before discussing the proposed framework, let's first understand the principle of the ISAC technique. The ISAC radar operates on the Matched Filter Fast Fourier (MF-F) algorithm, which enhances the estimation of detection parameters in radar ISAC. The MF-F algorithm offers several advantages, such as improved precision through the use of the Fibonacci sequence, fractional precision estimates of detection parameters, reduced number of comparisons required, and low complexity compared to other estimation algorithms. It has also shown resilience in numerical simulations, ensuring accurate estimates even in challenging conditions. This algorithm is used for various parameters estimation and data detection. The Doppler estimation, supported by ISAC radar, is improved through an iterative refinement process, ensuring higher accuracy and better data detection. The choice between OFDM and OTFS greatly impacts system performance. Once the user's parametersare estimated, the parameters value can be determined. The receiver processing enables the implementation of two modes for ISAC radar applications:

active detection and joint passive detection. The objectives of these two modes are detailed as follows [30], [31]:

- The objective of active detection is to calculate channel delay $\tau$ and Doppler shift $v$ by considering transmit vector $X$ and receive vector.
- The primary goal of passive detection and joint data detection is to estimate the channel parameters $(\alpha, \tau, v)$ and retrieve both $X$ and the received vector $Y$.

To obtain the parameters of the i[th] target, we need to consider the useful signal, represented by $\alpha_i \Gamma_i s$, and the interference signal, represented by $\sum_{j \neq i} \alpha_j \Gamma_j s$. In order to effectively estimate the parameters, it is crucial to implement an interference cancellation mechanism.
This mechanism involves eliminating the interference signal from the previously estimated (i-1) targets. Specifically, for the first target with parameters $(\alpha_1, \tau_1, v_1)$, the estimator is provided as follows [30], [31]:

$$(\hat{\alpha}_1, \hat{\tau}_1, \hat{v}_1) = \arg max_{(\hat{\alpha}_1, \hat{\tau}_1, \hat{v}_1) \epsilon \Lambda_i} \|(\Gamma_i)^H - y\|^2 \qquad (20)$$

where $(\hat{\alpha}_1, \hat{\tau}_1, \hat{v}_1)$ are target parameters the estimated Doppler and $c_0$ the velocity of sight.
The estimated delay and Doppler can be presented firstly based on the indication expression [30], [31] as follows:

$$(\hat{\tau}_i, \hat{v}_j) = \arg max_{(\tau,v) \epsilon \Lambda_i} |(\Gamma_i)^H Y_i|^2 \qquad (21)$$

Where $\Gamma_i$ delay Doppler plane, $\hat{v}_i$ is the estimated Doppler and $c_0$ the velocity of sight and $f_c$ the carrier frequency. We can present also, the estimated velocity is indicated as follows[30], [31]:

$$\hat{v}_i = \frac{\hat{v}_i c_0}{2 f_c} \qquad (22)$$

where $\hat{v}_i$ is the estimated Doppler and $c_0$ the velocity of sight?
Once the estimated delay and Doppler are obtained, the target range can be calculated using the following equation from [30], [31], [32]:

$$\hat{r}_i = \frac{\hat{\alpha}_i c_0}{2} \qquad (23)$$

In this study, we delve deeper into examining the precision of OTFS-OFDM sensing through the usage of a novel two phases estimation technique that integrates FFT and SFFT. The evaluation criterion employed is the root mean square error (RMSE). The RMSE value for the range metric is denoted as [30], [31], [32]:

$$RMSE(r) = \sqrt{\mathbb{E}_{r \epsilon \mathcal{D}} \left[ \frac{1}{P} \sum_{i=1}^{\mathcal{D}} (r_i - \hat{r}_i)^2 \right]} \qquad (24)$$

Where the Real values, are denoted by $r = [r_1, \cdots, r_P]$, and the estimated values, are denoted as $\hat{r} = [\hat{r}_1, \cdots, \hat{r}_P]$, represent the target range. The simulated samples set is $\mathcal{D}$. Based on the estimated parameters, we propose a dynamic waveform selection strategy that empowers operators to optimally choose the most suitable waveform for mobile users. This decision is made by comparing a predetermined threshold parameter with the estimated ones. The proposed architecture enhances the performance of wireless communication systems in various scenarios of mobility levels by selecting the OTFS strategy for high mobility scenarios where the estimated performance metric surpasses a specific threshold ($\hat{P} \geq P_{\text{threshold}}$), and the OFDM strategy for low mobility scenarios where the estimated performance metric falls below the threshold value ($\hat{P} < P_{\text{threshold}}$), where $P_{\text{threshold}} = \{range = 30m, velocity = 120 km/h\}$)[1],[32].

## 5. RESULTS AND DISCUSSIONS

To evaluate the aforementioned approach, simulations were conducted under the conditions summarized in Table 2 below.

Table 2. Simulation Parameters of OTFS and OFDM.

| Parameter | Value |
|---|---|
| Channel Power Delay Profile | EVA |
| Subcarrier Spacing $\Delta f$ | 15 KHz |
| Number of symbols per frame | 128 |
| Number of Subcarriers per Block | 16 |
| Carrier Frequency $f_c$(GHz) | 0.95 |
| Velocity V(km/h) | 3,10,200,500 |
| Modulation | 64QAM |

### 5.1. Complexity Analysis

In this section, we discuss the complexity of the proposed hybrid system that switches between OTFS and OFDM waveforms in the transmitter and receiver signal processing chains. We provide a detailed analysis of the complexity of the MF-F algorithm, which is divided into two parts. The first part, the MF step, involves a low complexity circular shift operation with a complexity of $\mathcal{O}(M) + \mathcal{O}(N)$, resulting in a total complexity of $\mathcal{O}(MN)$, and a multiplication operation. The second part, step F, involves matrix calculations with a complexity of $\mathcal{O}(MN)^2$ when performed directly. The multiplication of the diagonal matrix has a complexity of $\mathcal{O}(MN)$, and the cyclic shift matrix operation has a complexity of $(\mathcal{O}(MN)^2)$.

The FFT at point M and inverse FFT at point N have respective complexities of $\mathcal{O}((MN)\log_2(M))$ and $\mathcal{O}((MN)\log_2(N))$. Therefore, the total complexity of the proposed MF-F algorithm is of the order of $\mathcal{O}((MN)\log_2(MN))$. The Table 3 summarizes the various complexity parameters of various algorithms used in different waveforms.

Table 3. Complexity Analysis

| Waveform's | Algorithm | Complexity |
|---|---|---|
| OFDM [18] | FFT | $\mathcal{O}((N)\log(N))$ |
| OTFS [26], [27], [28] | SFFT | $\mathcal{O}((MN)\log_2(MN))$ |
| Proposed | FFT(if $\check{p} \leq p_{threshold}$) | $\mathcal{O}((N)\log(N))$ |
|  | SFFT(if $\check{p} > p_{threshold}$) | $\mathcal{O}((MN)\log_2(MN))$ |

The aim of deploying the ISAC radar system for OTFS-OFDM is to enhance the performance of the algorithm in practical scenarios. The integration of the ISAC radar significantly reduces the complexity associated with the implementation of OFDM/OTFS. Our proposed architecture, designed to facilitate a guided switch between OTFS and OFDM, ensures high data detection and low implementation complexity, particularly in high mobility scenarios, due to MF-F algorithm. This approach not only improves system efficiency but also reduces complexity. Furthermore, the integrated approach enhances system efficiency, allows for adaptation to changing channel conditions, and improves the robustness of the communication system. It also enhances data quality, robustness, and mobility. The proposed framework offers a comprehensive and global solution to address the complexity of OTFS and the challenges faced during ahigh mobility for OFDM. The integration of ISAC radar simplifies the implementation of OTFS-OFDM, guaranteeing high data detection and low implementation complexity. This is especially advantageous in rapid moving cases. Overall, the proposed architecture improves system efficiency, adapts to changing channel conditions, and enhances the robustness of the communication system, thereby improving data quality, robustness, and mobility.

### 5.2. Evaluating Performance of Radar ISAC System

The simulation involved conducting 200 Monte Carlo experiments for each algorithm using three sets of estimated parameters, namely {($R_1$ = 10 m, $V_1$ = 3km/h, $V_2$ =10km/h, $V_3$ = 30km/h, $V_4$= 200km/h, $V_5$ = 500km/h)}, {($R_2$ = 30 m, $V_1$ = 3km/h, $V_2$ =10km/h, $V_3$ = 30km/h, $V_4$= 200km/h, $V_5$ = 500km/h)}, and {($R_3$ = 90 m, $V_1$ = 3km/h, $V_2$ =10km/h, $V_3$ = 30km/h, $V_4$= 200km/h, $V_5$ = 500km/h)}.The results presented in Fig.4 indicate that the proposed algorithm in this study outperforms other algorithms in terms of Range Sensing performance.

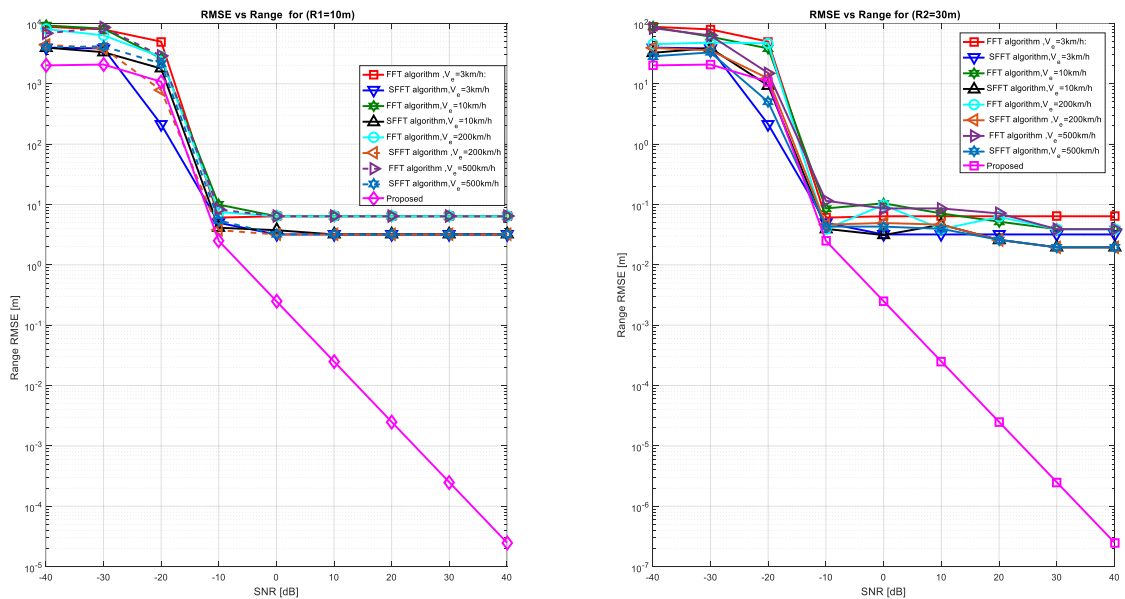

(a)  (b)

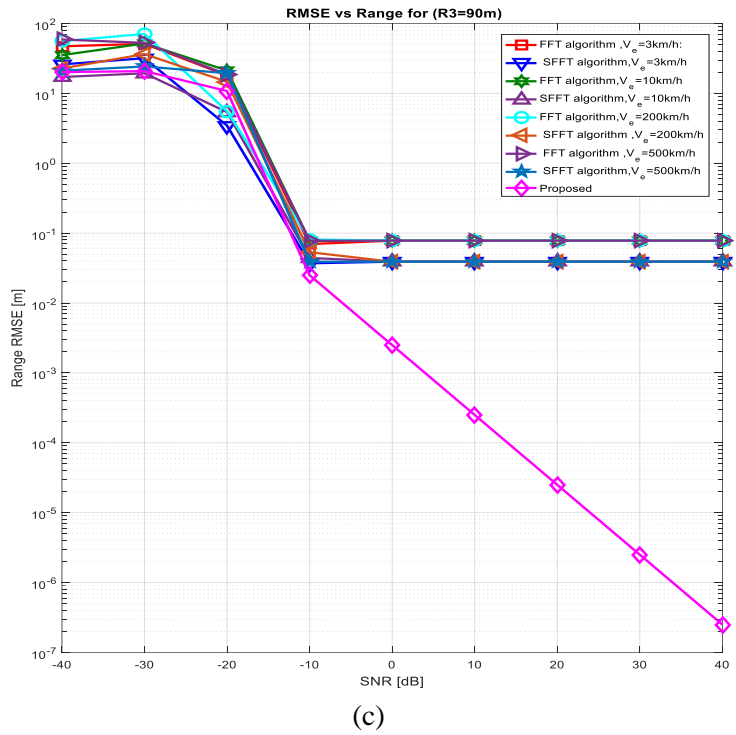

(c)

Fig. 4. RMSE vs. range under various algorithm.

The simulation results presented in Fig.4 showcase the suboptimal Range-sensing performance of the FFT algorithm compared to the other algorithms. Nevertheless, the FFT and SFFT algorithms present a comparable Range sensing performance. In addition, Fig.4 illustrates the range sensing RMSE profile at SNR=10 dB, where the suggested algorithm that combines FFT and SFFT. This exceedsother algorithms dealing with perceptual accuracy, with lower RMSE values. Furthermore, the proposed algorithm demonstrates robustness under varying conditions, with stable perceptual RMSE values, while other algorithms exhibit instability and lack of robustness. In Fig.5, a comparison between our proposed approach and existing techniques, namely FFT, MF-F, and SFFT is presented, in terms of velocity estimation versus SNR radar.

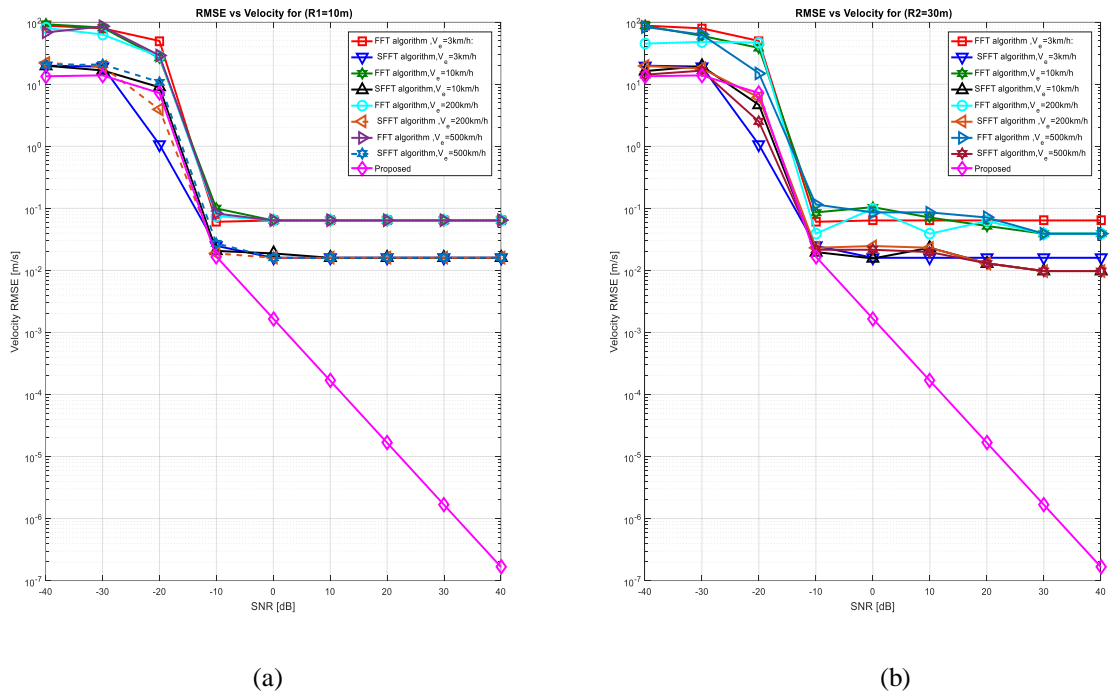

(a)                                                                                   (b)

Fig.5. RMSE vs. Velocity under various algorithm.

Figs 5(a) and 5(b) illustrate that our proposed framework achieves a substantial improvement in resolution using the same resources, highlighting its superiority in velocity estimation compared to existing methods. In brief, the results demonstrate that our suggested architecture outperforms current techniques in velocity estimation. The significant enhancement in resolution achieved by our technique can greatly enhance object detection and tracking, which has significant implications for various applications in wireless communication. To conclude, the combined FFT and SFFT algorithm provides a versatile and robust approach for radar detection, demonstrating improved performance in various conditions. Its adaptability and efficiency make it a valuable tool in radar technology. The results suggest that our ISAC system for speed estimation exhibits exceptional performance across multiple metrics, with high accuracy and robustness under challenging conditions.

### 5.3. BER Performance

In this section, the Bit Error Rate (BER) performance of the proposed OTFS-OFDM method is evaluated for different Doppler frequencies. Initially, an OFDM system is considered as a function of estimated speed, and the corresponding BER is depicted in Fig.6 for different Doppler frequencies. The modulation scheme employed is 64QAM.

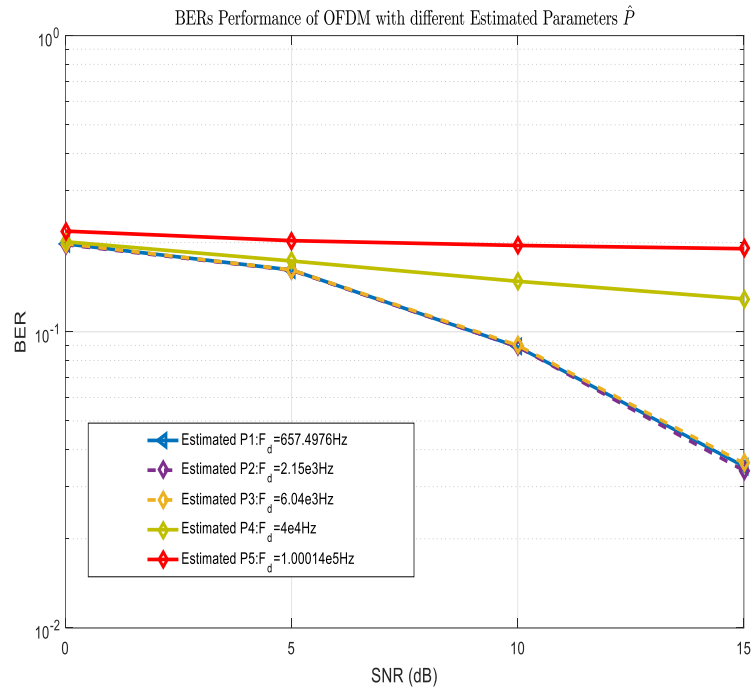

Fig.6. BERs Performance of OFDM

In Fig.6, we have selected five different values for the Doppler Frequency estimate: $\hat{F}_d$=657.4976Hz, $\hat{F}_d$ =2.15 kHz, $\hat{F}_d$ =6.04 kHz, $\hat{F}_d$ =40 kHz, and $\hat{F}_d$ =114 kHz. When the estimated speed is low (3km/h, 10km/h, and 30km/h), the BER values are similar for all cases. However, as the Doppler estimate increases to $\hat{F}_d$ =40 kHz and $\hat{F}_d$ =114 kHz, the BER turns highest among these values. This suggests that higher Doppler frequencies lead to an increase in BER due to inaccurate channel estimation and failure in data detection caused by the allocation of signal power to the channel speed. These findings indicate that as mobility speed increases, OFDM users experience a decline in BER performance. This is because OTFS applies high mobility to achieve DD diversity, while the orthogonality of OFDM subcarriers is compromised in highly mobile scenarios. Fig. 6 demonstrates that the Higher order Modulation scheme requires a big SNR levels to attain a good BER.The BER is assessed from 0 dB to 15 dB for a 64-QAM modulation scheme. It is noticed that OFDM at low Doppler frequency yields the best BER, while OFDM at high mobility results in the lowest BER values within the 0 dB to 15 dB range. Therefore, using low mobility for OFDM processing aligns well with the simulation outcomes in Fig. 6. The findings demonstrate that in scenarios with low mobility, the OFDM approach offers nearly optimal BER that is not affected by the speed utilized. However, as the Doppler frequency increases, the BER encounters a significant rise, leading to suboptimal performance. It is crucial to reconsider the choice of processing tools to address the drawbacks of OFDM in such scenarios. To deal with the problem of high mobility of the waveform, we propose the OTFS waveform as the optimal solution for the proposed waveform. The SNR and the estimation of various parameters assisted by ISAC radar are less affected by the OTFS waveform. Therefore, the higher the noise level, the more important it is to use an ISAC radar for accurate speed estimation.

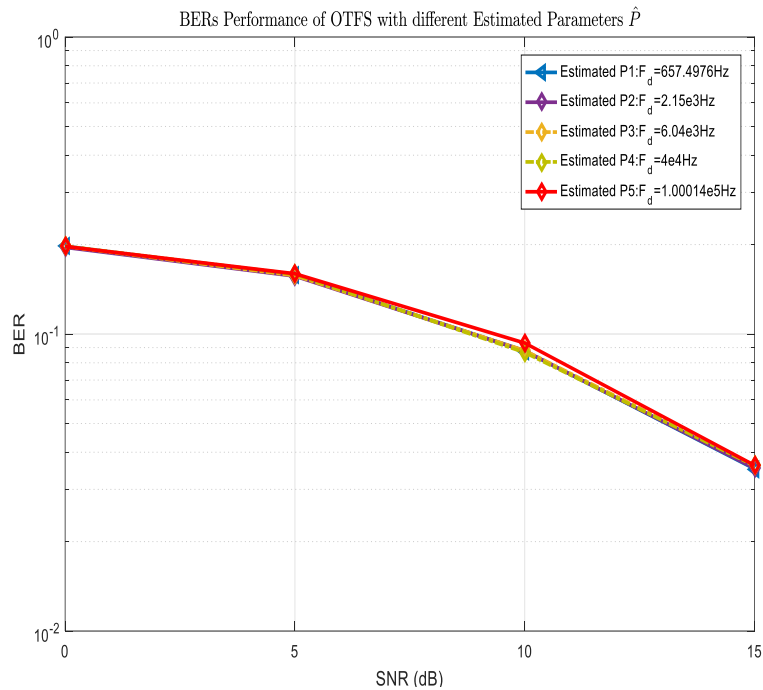

Fig.7. BERs Performance of OTFS.

Fig.7 provides a comprehensive illustration of the intricate relationship between Bit Error Rate (BER), Signal-to-Noise Ratio (SNR), and estimated Doppler frequency in the context of 64-QAM modulation scheme. The figure elucidates that, for a given SNR value, the BER exhibits a positive correlation with the estimated Doppler frequency, a phenomenon commonly observed in communication systems operating at various speeds. Furthermore, the figure underscores the impact of increased speed on system performance, as manifested by an elevated BER. Remarkably, both low and high rate Orthogonal Time Frequency Space (OTFS) schemes demonstrate superior BER performance compared to other approaches. This observation can be attributed to the unique characteristics of OTFS, which enable it to effectively mitigate the deleterious effects of high Doppler frequencies and mobility. Moreover, Figure 7 reveals a salient aspect of higher order modulation schemes, namely their requirement for a higher SNR to attain a favourable BER. The BER assessment, conducted across an SNR range of 0 dB to 15 dB, corroborates this assertion. Notably, OTFS at low Doppler frequencies consistently offers the best BER performance. However, at high mobility, the results are nearly identical, yielding the lowest BER values within the 0 dB to 15 dB range. The findings presented in Figure 7 substantiate the efficacy of employing high mobility for OTFS processing, aligning with the simulation outcomes. The results unequivocally demonstrate that in high-mobility scenarios, the OTFS approach delivers nearly optimal BER performance, unaffected by the high speeds utilized. This underscores the robustness and versatility of the OTFS scheme in handling diverse mobility conditions.Thereby offering a promising evolution for future research and development in the realm of wireless communication.

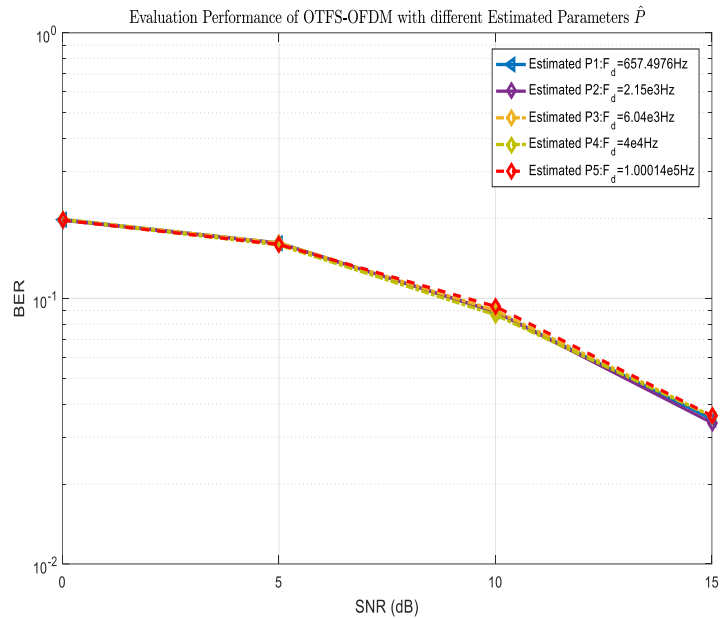

Fig.8. BERs Performance of OTFS-OFDM

The Fig.8 presents the BER performance of a hybrid scheme for OTFS-OFDM systems, employing a 64-QAM modulation scheme. This hybrid scheme combines the advantages of both waveforms to raise BER performance in rapidly movement. By applying OFDM processing in parallel, the issue of high mobility is effectively addressed. The OTFS filter proves to be superior due to its ability to reduce noise caused by the speed of the moving object. These observations strongly support the selection of OTFS in high-speed scenarios. However, when considering low Doppler frequencies, the processing complexity of OTFS compared to OFDM favors the retention of OFDM due to its simpler implementation. This can be attributed to the Doppler effect, which induces changes in signal frequency and phase at higher speeds. The figures clearly show the suitability of OTFS, as no performance degradation is observed under such conditions. These results further reinforce the preference for OTFS in high mobility scenarios. Moreover, the simulation results in Figure 6 align well with the use of less OFDM processing, effectively resolving the high mobility problem. Even though this technique can be utilized for lower mobility, its processing complexity is higher compared to that of OFDM. Despite its potential for low mobility applications, the processing complexity of OTFS often results in OFDM being preferred due to its simpler implementation. An interesting observation is that the BER curves for all scenarios converge, indicating that user mobility has little impact. This highlights the significance of OTFS in high mobility scenarios. It is worth noting that the data reveals that a highly mobile user achieves a BER almost identical to that of a user with low mobility. The intriguing observation is that the BER curves for all scenarios converge, indicating that user mobility has little impact. This characteristic highlights the significance of OTFS in faster mobility. It is worth noting that the data reveals that a highly mobile user achieves a BER that is nearly identical to that of a user with low mobility. Additionally, this technique can be applied to low speeds, although its processing complexity is higher compared to OFDM. In spite of the potential application at low speeds, the complexity of implementing OTFS often drives to the preference of OFDM due to its simpler implementation. Another interesting consequence is that both OTFS and OFDM waveforms exhibit significant variations in BER, suggesting that the choice of waveform can be guided by radar ISAC based on parameter's estimation such as Doppler and speed estimation. Specifically, this approach can enhance BER for high mobility users when utilizing the OTFS waveform. Conversely, for low mobility users, the OFDM waveform seems to be a reliable choice. The results demonstrate that the hybrid scheme outperforms the employment of OTFS or OFDM waveforms alone. Lower mobility cases, the BER of the hybrid scheme is similar to that of using the OFDM waveform alone, as the Doppler spread can be disregarded, making the use of OTFS waveform less advantageous. However, it is crucial to note that the performance of the hybrid scheme heavily relies on selecting appropriate parameters, such as the subcarrier spacing and the delay Doppler grid size of the OTFS. Furthermore, implementing a hybrid scheme is more complex than using either OTFS or OFDM waveforms separately, which may be a concern in certain practical applications. The findings demonstrate that the hybrid scheme outperforms the usage of OTFS or OFDM waveforms individually. In cases with low mobility, the BER of the hybrid scheme is comparable to that of the OFDM waveform, as the impact of Doppler spread can be disregarded, making the OTFS waveform less advantageous. However, it is important to emphasize that the performance of the hybrid scheme heavily depends on the selection of appropriate parameters, such as the subcarrier spacing and the delay Doppler grid size of the OTFS. Additionally, implementing the hybrid scheme is more complex compared to using OTFS or OFDM waveforms separately, which may raise concerns in practical applications. In conclusion, the results suggest that the hybrid scheme is a viable option for enhancing BER performance in higher mobility situations.

## 6. CONCLUSION

This paper has presented a novel concept for a hybrid system that enables the selection of either OTFS or OFDM waveforms by alternating between signal processing chains in the transmitter and receiver. The system is based on the ISAC system assistance, which estimates and incorporates delay, Doppler, velocity, and range for sensing and communication purposes. The focus of our research is on OTFS, a waveform that is gaining popularity due to its availability in high user mobility scenarios. Our proposed strategy aims to make a validated selection between OTFS and OFDM to better address user mobility needs. The ISAC system further improves the selected approach by providing accurate estimation, low complexity, and high-resolution range. Our study has resulted in a more suitable and useful processing approach based on the user's mobility level. The findings of our research have validated the effectiveness of our proposed concept and its sustainability in really livedworld systems with switching procedures. We suggest the possibility of developing more advanced strategies in the future, such as adaptive or automated switching procedures based on different criteria for immediate implementation. In conclusion, the switching selection strategy acts as a powerful tool for enhancing the performance of OTFS-OFDM systems, paving the way for promising results and empowering rich future research and development.

### CONFLICTS OF INTEREST

The authors declare no conflict of interest.

### REFERENCES

[1]  Wyche, and al., " Coexistence Analysis of OTFS and OFDM Waveforms for Multi mobility Scenarios, " 95th Vehicular Technology Conference IEEE, 2022.
[2]  B. Mohammed, and al. " 6G mobile communication technology: Requirements, targets, applications, challenges, advantages, and opportunities, " Alexandria Engineering Journal, 2022.
[3]  S. Usha, and al. " A New Approach to Improve the Performance of OFDM Signal for 6G Communication." Int. J. Compute. Networks Common, 2022.
[4]  A. Rajendrasingh and al., " OFDM systems resource allocation using multi-objective particle swarm optimization." Int. J. Compute. Networks Common, 2012

International Journal of Computer Networks & Communications (IJCNC) Vol.16, No.4, July 2024

[5]    Z. Wei, and al. "Orthogonal Time Frequency Space Modulation: A promising next generation waveform, " IEEE Wireless Common., vol. 28, pp. 136–144, 2021.
[6]    Zhengyuan, et al. "Toward 6G Multicell orthogonal time frequency space Systems: Interference Coordination and Cooperative Communications." IEEE Vehicular Technology Magazine ,2024
[7]    LaGuardia and al., " OTFS vs. OFDM in the presence of sparsity: A fair comparison, " IEEE Transactions on Wireless Communications, 2021.
[8]    Napata, and al. " Joint Compensation of TX/RX IQ Imbalance and Channel parameters for OTFS Systems under Timing Offset. " National Conference on Communications (NCC). IEEE, 2023.
[9]    Damián, and al. "Equalizers Performance Enhancing in MISO-OTFS Configuration." IEEE International Workshop on Mechatronic Systems Supervision (IW_MSS). p. 1-6, 2023.
[10]   Zwinglian, and al. "A survey on high mobility wireless communications: Challenges, opportunities and solutions. " IEEE Access, 2016.
[11]   Hadani, and al., " Orthogonal time frequency space modulation, " IEEE Wireless Communications and Networking Conference (WCNC), 2017.
[12]   Hay, and al. " Delay Doppler Communications: Principles and Applications." Elsevier, 2022.
[13]   M. Abderrahim, et al. "Performance Evaluation of OTFS and OFDM for 6G Waveform.", ITM Web of Conferences, 2022.
[14]   Horiuchi, and al. " Channel Estimation and Equalization for CP-OFDM based OTFS in fractional Doppler channels. " IEEE International Conference on Communications Workshops (ICC Workshops). IEEE, 2021.
[15]   Laotian, and al. " Doubly Iterative Sportified MMSE Turbo Equalization for OTFS Modulation. « IEEE transactions on communications, 2023.
[16]   Sneha, and al. " Analysis of PAPR in OTFS Modulation with Classical Selected Mapping Technique. " International Conference on Communication Systems & Networks IEEE, 2023.
[17]   Shi, Jia, and al. " OTFS enabled LEO Satellite Communications: A Promising Solution to Severe Doppler Effects. " IEEE Network, 2023.
[18]   Lorenzo, and al. " Performance analysis of joint radar and communication using OFDM and OTFS. " 2019 IEEE International Conference on Communications Workshops. IEEE, 2019.
[19]   Zhou, and al. " Learning to equalize OTFS. ", IEEE Transactions on Wireless Communications, 2022
[20]   S. Yoke, and al., " Real time 2D velocity localization measurement of a simultaneous-transmit OFDM MIMO Radar using Software Defined Radios ", European Radar Conference (Eura), IEEE, 2016.
[21]   P. John and al. " Digital communication through band limited channels. "Digital Communications, 2008.
[22]   T. Danny and al. " Integrated sensing and communication in 6G: Motivations, use cases, requirements, challenges and future directions, " IEEE International Online Symposium on Joint Communications and Sensing, IEEE, 2021.
[23]   Maksim, and al., "  Integration of communication and sensing in 6G: A joint industrial and academic perspective, " IEEE 32nd Annul. Int. Symp. Pers., Indoor Mobile Radio Commun. , PIMRC, 2021.
[24]   F. Liu, and al," Integrated sensing and communications: Toward dual-functional wireless networks for 6G and beyond," IEEE J. Sel. Areas Commun., 2022.
[25]   T. Wild, and al.," Joint design of communication and sensing for beyond 5G and 6G systems. ", IEEE Access, 2021.
[26]   Y.Aktan, and al., " Haar-MUSIC: A New Hybrid Method for Frequency Estimation. " Circuits, Systems, and Signal Processing ,2023.
[27]   P. Yuan, and al., " A Hybrid Frame Structure Design of OTFS for Multi-tasks Communications,"IEEE 34th Annual International Symposium on Personal, Indoor and Mobile Radio Communications (PIMRC), Toronto, ON, Canada, 2023.
[28]   Z.Tomasz., and al. "Wireless OTFS-based Integrated Sensing and Communication for Moving Vehicle Detection." IEEE Sensors Journal , 2024.
[29]   X.Zichao, and al. "A novel joint angle range velocity estimation method for MIMO-OFDM ISAC systems." arXiv preprint arXiv: 2308.03387, 2023.
[30]   T.Zhiling, and al. " The Estimation Method of Sensing Parameters Based on OTFS. " IEEE Access,2023

[31] W.Yongzhi, and al. " DFT-spread orthogonal time frequency space system with superimposed pilots for terahertz integrated sensing and communication.," IEEE Transactions on Wireless Communications,2023.
[32] Z.Salah Eddine, and al. "OTFS-Based ISAC for Super-Resolution Range-Velocity Profile ". IEEE Transactions on Communications, 2024

**AUTHORS**

**Amina Darghouthi** was born in Tozeur Tunisia, in 1993. Doctoral student researcher in electrical engineering at the National Engineering School of Gabes (Tunisia).She is an esteemed member of the Research Laboratory Modeling, Analysis, and Control Systems (MACS), registered under LR16ES22 (www.macs.tn), actively involved in research. In addition to her research pursuits, Fatma is currently serving as a contractual lecturer at the National School of Engineers of Gabes, where she shares her knowledge and expertise with students.

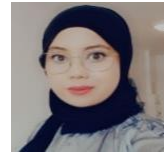

**Abdelhakim Khlifi**, Innov'COM laboratory, Sup'COM, University of Carthage, Tunisia

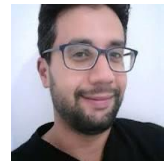

**Belgacem Chibani**, 3MACS Laboratory: Modelling, Analysis and Control of Systems, University of Gabes, Tunisia

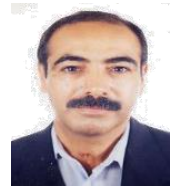